\documentclass{article}
\usepackage{spconf,amsmath,graphicx,hyperref}

\usepackage[T1]{fontenc}
\usepackage[utf8]{inputenc}
\usepackage{cite}
\usepackage{amssymb,amsfonts}
\usepackage{algorithmic}
\usepackage{textcomp}
\usepackage{xcolor}
\usepackage{tabularx}
\def\BibTeX{{\rm B\kern-.05em{\sc i\kern-.025em b}\kern-.08em
    T\kern-.1667em\lower.7ex\hbox{E}\kern-.125emX}}

\makeatletter
\def\thebibliography#1{%
  \section{References}%
  \footnotesize
  \list{[\arabic{enumi}]}{%
    \settowidth\labelwidth{[#1]}%
    \leftmargin\labelwidth
    \advance\leftmargin\labelsep
    \usecounter{enumi}%
    \setlength{\itemsep}{0pt}%
    \setlength{\parsep}{0pt}%
    \setlength{\parskip}{0pt}%
    \setlength{\topsep}{0pt}%
  }%
  \def\newblock{\hskip .11em plus .33em minus .07em}%
  \sloppy
  \clubpenalty4000
  \widowpenalty4000
  \sfcode`\.=1000\relax
}

\makeatother

\title{Influence of Geometrical Acoustic Simulator Complexity\\on a Trained Multisource Localizer}

\name{%
\begin{tabular}{@{}c@{}}
Fabian Staub$^{\star\dagger}$, Nils Meyer-Kahlen$^{\star}$, Thomas Deppisch$^{\star}$, Sergio de las Heras$^{\star}$,\\
\textit{Florian Klein}$^{\ddagger}$, \textit{Stephan Werner}$^{\dagger}$, and \textit{Johannes M. Arend}$^{\star}$
\end{tabular}}
  
\address{$^{\star}$ Department of Information and Communications Engineering, Aalto University, Espoo, Finland \\
         $^{\dagger}$ Institute for Media Technology, Technische Universität Ilmenau, Ilmenau, Germany \\
         $^{\ddagger}$ Fraunhofer Institute for Digital Media Technology, Ilmenau, Germany}

\begin{document}
\ninept
\maketitle

\begin{abstract}

Learning-based acoustic sound source localization and detection (SSLD) requires large labeled datasets covering diverse acoustic conditions. Since obtaining measured data is costly, training commonly uses simulated data, while practical devices must operate under real-world conditions. However, higher simulation complexity increases data-generation cost, and the complexity required for reliable generalization remains unclear. In practice, SSLD methods often rely on efficient geometrical acoustic simulation, typically the image-source method. This work investigates how geometrical acoustic simulation complexity affects the real-world performance of a common multi-source SSLD model. We train the model with simulators ranging from anechoic conditions to high-order image-source simulations, optionally including diffuse reverberation, array simulation, and randomized image-source positions, and evaluate them on three measurement-based test datasets. Results show that anechoic simulation is insufficient, while medium-complexity image-source simulations already provide strong real-world performance. Further increases in complexity yield only marginal gains, with the best performance obtained using the highest tested image-source order, array simulation, and randomized image-source positions. For this configuration, measured-domain performance approaches within-domain simulated performance, suggesting limited benefit from further increasing simulation complexity. These findings show how the complexity--performance trade-off can be exploited in future data-driven SSLD, and highlight image-source randomization as an efficient way to improve generalization.
\end{abstract}

\section{Introduction}\label{sec:intro}
In recent years, many classical audio signal processing tasks, including sound source localization and detection (SSLD), have been increasingly tackled using deep learning methods. This change has shifted the focus of engineers and researchers from deriving parameter estimators toward designing and curating suitable training datasets. For acoustic problems, however, relying solely on measured real-world data is often prohibitively labor-intensive. Acoustic simulators offer a practical alternative, provided that training on the resulting data generalizes well to real-world conditions. Due to their low computational cost, simulators based on geometrical acoustics, in particular Python implementations~\cite{scheiblerPyroomacousticsPythonPackage2018} of the image-source method (ISM)~\cite{allenImageMethodEfficiently1979}, are commonly employed~\cite{saviojaOverviewGeometricalRoom2015}.

Despite their widespread use, simulator design choices have received little attention. Addressing this gap is important because practical applications often require balancing simulation cost against improvements in real-world generalization. Recently, studies have indicated an advantage of wave-based acoustic simulators over geometric simulators in speech enhancement and dereverberation tasks~\cite{gotz_improving_2026, gotz_room-acoustic_2025, milo_training_2026}. However, generating large-scale datasets with wave-based methods remains computationally very expensive. At the same time, the cost-performance trade-off of more practical geometrical simulators and their effect on SSLD performance have not yet been explored in full detail. 

So far, only~\cite{Gelderblom2021a,srivastava_how_2023} have examined how the complexity of geometrical acoustic simulation affects learning-based SSLD performance. However, these studies focus on factors such as source and receiver directivity, diffuse reflections, and frequency-dependent absorption, leaving factors such as image-source order, dedicated late-reverb synthesis, or microphone array transfer functions (ATF) open, which may involve important cost-benefit trade-offs.

In this work, we train a multi-source SSLD model~\cite{poschadelInvestigationsHigherorderSpherical2025} on different simulated datasets. These datasets cover conditions ranging from anechoic environments to different image-source orders, optionally including diffuse late reverberation, measured ATFs for array simulation, and randomized image-source positions, similar to \cite{Luo2024}. We then evaluate the models on three real-world, measured datasets and compare their performance with the Multiple Signal Classification (MUSIC) algorithm~\cite{Schmidt1986}, which is provided with the number of active sources and therefore serves only as a reference for localization performance. Additionally, we discuss the computational cost of the different simulators and its relation to SSLD performance. To foster reproducible research, code for data generation, simulation, training, and evaluation is available online\footnote{\url{https://github.com/Fabian-Staub/acoustic-simulation-complexity-ssld}}.

Our results show that increased simulation complexity generally improves real-world SSLD performance, but that medium-complexity geometrical simulation already provides strong generalization, and that randomizing image-source positions yields a slight additional improvement. Together, these findings motivate effective low-cost simulation strategies for future SSLD systems.

\section{Datasets}\label{sec:datasets}
The following describes both the simulation-based training datasets and the measurement-based test datasets. As detailed in Sec.~\ref{sec:ssld_model}, the employed SSLD model uses third-order Ambisonic~\cite{rafaelyFundamentalsSphericalArray2019, zotter2019ambisonics} input features. Consequently, both the late reverberation and the microphone array simulation are represented in the spherical harmonic (SH) domain.

\subsection{Simulated Datasets}
We simulated 20 training dataset configurations to study the influence of four main simulation factors: ISM reflection order, diffuse late reverberation, image-source randomization, and microphone-array simulation. The full set of configurations is summarized in the top row of Fig.~\ref{fig:overview_results_mic_order_late_random}, which also shows the corresponding evaluation results. Each dataset was split at the room level into training, validation, and test sets using a 70/20/10 split. The held-out test split was used for within-domain evaluation.

\subsubsection{Geometrical Acoustic Simulation}
As a base, we used the ISM implementation from Pyroomacoustics~\cite{scheiblerPyroomacousticsPythonPackage2018}, which is among the most widely used geometrical acoustic simulators for training acoustic machine learning models, see e.g., ~\cite{Grinstein2023, roman2024spatialscaperlibrarysimulate, HuJinbo2024SMwE}. We generated datasets of 2,000 rooms with randomly sampled, constrained dimensions, each containing eight receiver positions and three sources (see Tab.~\ref{tab:room_configuration}). For each room, a reverberation time (RT) profile was randomly drawn from the measured impulse responses in~\cite{traerStatisticsNaturalReverberation2016} to resemble the RT distribution of real-world acoustic environments. Assuming equal absorption for all walls, octave-band absorption coefficients were derived from the room dimensions and the RT profiles using the inverse Eyring equation~\cite{eyringReverberation1930}. The datasets use ISM reflection orders of 0, 1, 2, and 12, corresponding to direct sound, low complexity, and high complexity simulations.

\subsubsection{Late Reverberation}\label{sec:late_reverb}
In selected datasets, diffuse late reverberation was added to the ISM simulations as uncorrelated, equal-energy noise across all Ambisonic channels, producing an isotropic diffuse sound field~\cite{epainSphericalHarmonicSignal2016}. The diffuse noise decay rate was adjusted according to the room RT profile and cross-faded with the simulated spatial room impulse response (SRIR) at the perceptual mixing time $t_{\mathrm{m}}$, which was fixed at 3.5~ms for zeroth-order ISM simulations and estimated from the room volume $V$ for all other simulations using~\cite{lindauMixingTime2012}
\begin{equation}
t_{\mathrm{m}} =
\left(
0.0117\,\frac{V}{\mathrm{m}^3} + 50.1
\right)\mathrm{ms}.
\label{eq}
\end{equation}
The decaying noise was scaled to match the late image-source energy, estimated here using an additional 12th-order simulation. Future work could use a more efficient estimation method based on an analytical expression for the late energy.

\subsubsection{Randomized Image-Source Positions}
The ISM simulations were based on ideal shoebox geometries. To introduce irregularity in selected datasets, image-source positions (excluding direct sound) were randomly perturbed by adding independent offsets in the $x$, $y$, and $z$ directions, each drawn uniformly from $[-d_{\max}, d_{\max}]$. Perturbed image sources whose distance to the receiver became shorter than the direct source--receiver distance were discarded, ensuring that no reflected path arrived before the direct sound. Using $d_{\max}\in\{1,10\}\,\mathrm{m}$ enabled the investigation of a transition from geometrically determined image sources to effectively uniformly distributed random ones.

\subsubsection{Microphone Array Simulation} 
Two microphone simulation approaches were compared. The first approach modeled an ideal SH receiver following the HARP implementation~\cite{sainiHARPLargeScaleHigherOrder2025}, with coincident virtual microphones corresponding to SH-basis patterns. To mimic the regularization effects observed in measured data~\cite{shaybetAmbisonicsNetworksEffect2024}, the simulated Ambisonic SRIRs were filtered with order-dependent high-pass filters derived from the ratio of regularized to ideal Eigenmike radial-filter magnitude responses, where the regularized responses were obtained using Tikhonov regularization with a maximum gain of 15~dB~\cite{politisComparingModeledMeasurementbased2017}.

The second approach directly simulated an em32 Eigenmike (EM) using measured ATFs from~\cite{politis2016microphone}. The resulting signals were encoded into Ambisonics using the discrete SH transform and radial filtering with the aforementioned regularization. By comparing these two approaches, we can assess whether explicitly simulating the characteristics of the real-world microphone improves performance on the measured test datasets.

\begin{table}[t]
    \centering
    \caption{Room Generation Setup}
    \label{tab:room_configuration}
    \setlength{\tabcolsep}{3pt}
    \begin{tabularx}{\columnwidth}{p{0.48\columnwidth} X}
        \hline
        \textbf{Parameter} & \textbf{Value} \\
        \hline
        Width/length & $[2.5,\,20.0]$~m, $0.5$~m steps \\
        Height & $[2.5,\,5.0]$~m, $0.5$~m steps \\
        Min. source--receiver dist. & $1.0$~m \\
        Min. source--surface dist. & $0.5$~m \\
        Min. receiver--surface dist. & $0.5$~m \\
        Number of rooms & 2,000 \\
        Receivers per room & 8 \\
        Sources per receiver & 3, separation $\geq 15^\circ$ \\
        \hline
    \end{tabularx}
\end{table}

\subsection{Measured Test Datasets}
\label{subsec:measured_datasets}

Three SRIR datasets measured with the em32 Eigenmike were used as test sets. All three were encoded into Ambisonics using the aforementioned discrete SH transform and Tikhonov-regularized radial filtering with a maximum gain of 15~dB.

%\subsubsection{Arni} 
\textit{Arni:}
The Arni dataset \cite{figueroa_duran_2025_15169199} was recorded in Aalto University's variable-acoustics room. It contains a total of 585 SRIRs from three Genelec 8010A sources at 65 receiver positions under dry, anisotropic, and reverberant conditions, with positions accurately measured using an OptiTrack system.

%\subsubsection{MOTUS} 
\textit{MOTUS:}
The MOTUS dataset~\cite{gotzDatasetHigherorderAmbisonic2021} contains 830 acoustic configurations of the same seminar room, measured with four Genelec 1030A sources at a single receiver position. We use sources 1--3 and exclude occluded-source measurements, resulting in 2436 SRIRs. Because the tape-measure source labels may be insufficiently precise, we estimate source directions from the direct-sound component of the multichannel SRIRs using the MUSIC algorithm. Validating this approach on the Arni dataset yielded a mean offset of 4.9° and a standard deviation of 2.0° relative to the OptiTrack labels.

%\subsubsection{AGrid} 
\textit{AGrid:}
The AGrid dataset is an unpublished dataset that was recorded in 14 different rooms at Aalto University, with two receiver positions and two source positions per receiver, yielding 56 SRIRs. Ground-truth directions of arrival (DOAs) are estimated using the same MUSIC-based approach as for the MOTUS dataset.

\subsection{Array Signals}
The SRIRs from the different datasets were convolved with TIMIT speech samples~\cite{garofoloTIMIT1993}, following the procedure in~\cite{poschadelInvestigationsHigherorderSpherical2025} with a few modifications. For each receiver position, audio samples were generated at a sampling rate of 16~kHz using three speaker tracks, one for each source SRIR. Each track contained two repetitions of a speech sample, with a random pause of 0--1.5~s inserted before, between, and after the repetitions. The mixture duration was determined by the longest speaker track and rounded up to the next full second. This yielded an approximately uniform distribution of frames containing between zero and three simultaneously active sources. Anechoic babble noise was convolved with a diffuse Ambisonic late-reverberation tail (see Sec.~\ref{sec:late_reverb}), whose octave-band decay times were obtained by averaging the RTs of the corresponding source SRIRs. The resulting reverberant babble noise was then added to the speech mixture at a signal-to-noise ratio of 20~dB. Finally, the mixture was split into one-second segments, which were used as model input.

\section{SSLD Model}\label{sec:ssld_model}
\subsection{Model Architecture} We adopted a convolutional recurrent neural network (CRNN) architecture, the most common choice for SSLD tasks, and based our model on~\cite{poschadelInvestigationsHigherorderSpherical2025}, which extends the SELDnet architecture~\cite{adavanneSoundEventLocalization2019,politisSTARSS22DATASETSPATIAL2022} to higher-order Ambisonic input features. The model used the activity-coupled Cartesian direction-of-arrival (ACCDOA) representation, which compactly encodes both source activity and direction~\cite{shimadaACCDOAActivityCoupledCartesian2021}.

\subsection{Input Features} 
As input features, we computed so-called higher-order pseudo-intensity vectors (HO-PIVs) and log-mel spectrograms from the $C=16$ channels of third-order ($N=3$) Ambisonic signals, as this was among the best-performing feature combinations in~\cite{poschadelInvestigationsHigherorderSpherical2025}. The HO-PIVs were computed using an STFT with a 640-sample Hann window, a 320-sample hop size, and a 1024-point FFT, resulting in $T=50$ frames and 513 frequency bins. To reduce feature dimensionality, the HO-PIVs were grouped and averaged into $F=128$ mel bands, and normalized as described in~\cite{poschadelInvestigationsHigherorderSpherical2025}.

The HO-PIVs were concatenated with 128-band log-mel magnitude spectra of all Ambisonic channels to improve source-number estimation. The resulting input had dimensions $(50 \times 128 \times 48)$, comprising 16 active HO-PIV, 16 reactive HO-PIV, and 16 magnitude-spectrum channels. HO-PIV and magnitude features were separately z-score normalized using training-set statistics.

\subsection{Training Parameters} We trained the model with Adam using an initial learning rate of $10^{-3}$ and weight decay of $3 \cdot 10^{-4}$, with a batch size of 800 and dropout rate 0.1. The learning rate was halved after five consecutive epochs without validation-loss improvement, and training was terminated after 20 consecutive epochs without validation-loss improvement. With these settings, training typically converged in about 100 epochs.

\section{Evaluation Metrics}\label{sec:metrics}
 
To assess both source localization and detection, we report localization error (LE) for localization accuracy, source-count mean absolute error (SC-MAE) for detection performance, and the $\mathcal{E}_{\mathrm{SSLD}}$ score as a combined measure. 

Active predictions are identified using an ACCDOA activity threshold of $\tau=0.5$ and matched to active ground-truth sources with the Hungarian algorithm~\cite{kuhnHungarianMethodAssignment1955}. The LE is then defined as the mean angular distance $d(\cdot,\cdot)$ over all true-positive (TP) prediction-reference pairs $(\hat{\mathbf{x}}_{kn},\mathbf{x}_{kn})$ across the $K$ evaluation samples,
\begin{equation}
\mathrm{LE}
=
\frac{
    \sum_{k=1}^{K}
    \sum_{n=1}^{\mathrm{TP}(k)}
    d\!\left(
        \hat{\mathbf{x}}_{kn},
        \mathbf{x}_{kn}
    \right)
}{
    \sum_{k=1}^{K}
    \mathrm{TP}(k)
},
\label{eq:le_all_samples}
\end{equation}
where $n$ indexes the matched TP prediction-reference pairs within evaluation sample $k$.

The SC-MAE measures detection performance as the mean absolute difference between the predicted ($\hat{N}_k$) and ground-truth ($N_k$) number of active sources,
\begin{equation}
\mathrm{SC\text{-}MAE}
=
\frac{1}{K}
\sum_{k=1}^{K}
\left|
\hat{N}_k - N_k
\right| \, .
\label{eq:sc_mae}
\end{equation}

As an overall metric for jointly evaluating localization and detection performance, we use the $\mathcal{E}_{\mathrm{SSLD}}$ score~\cite{poschadelInvestigationsHigherorderSpherical2025}, which combines the location-sensitive error rate $\mathrm{ER}_{15^\circ}$, location-sensitive false positive $\mathrm{F_{15^\circ}}$, the LE, and the localization recall (LR), 
\begin{equation}
\mathcal{E}_{\mathrm{SSLD}}
=
\frac{1}{4}
\left[
\mathrm{ER}_{15^\circ}
+
\left(1-\mathrm{F_{15^\circ}}\right)
+
\frac{\mathrm{LE}}{180^\circ}
+
\left(1-\mathrm{LR}\right)
\right] \, ,
\label{eq:essld}
\end{equation}
where lower values indicate better performance.

Statistical uncertainty in the evaluation metrics is estimated using the leave-one-out jackknife procedure described in~\cite{mesarosSoundEventDetection2019}, from which 95\% confidence intervals are estimated. 

\begin{figure*}[t]
    \centering
    \includegraphics[width=1\linewidth]{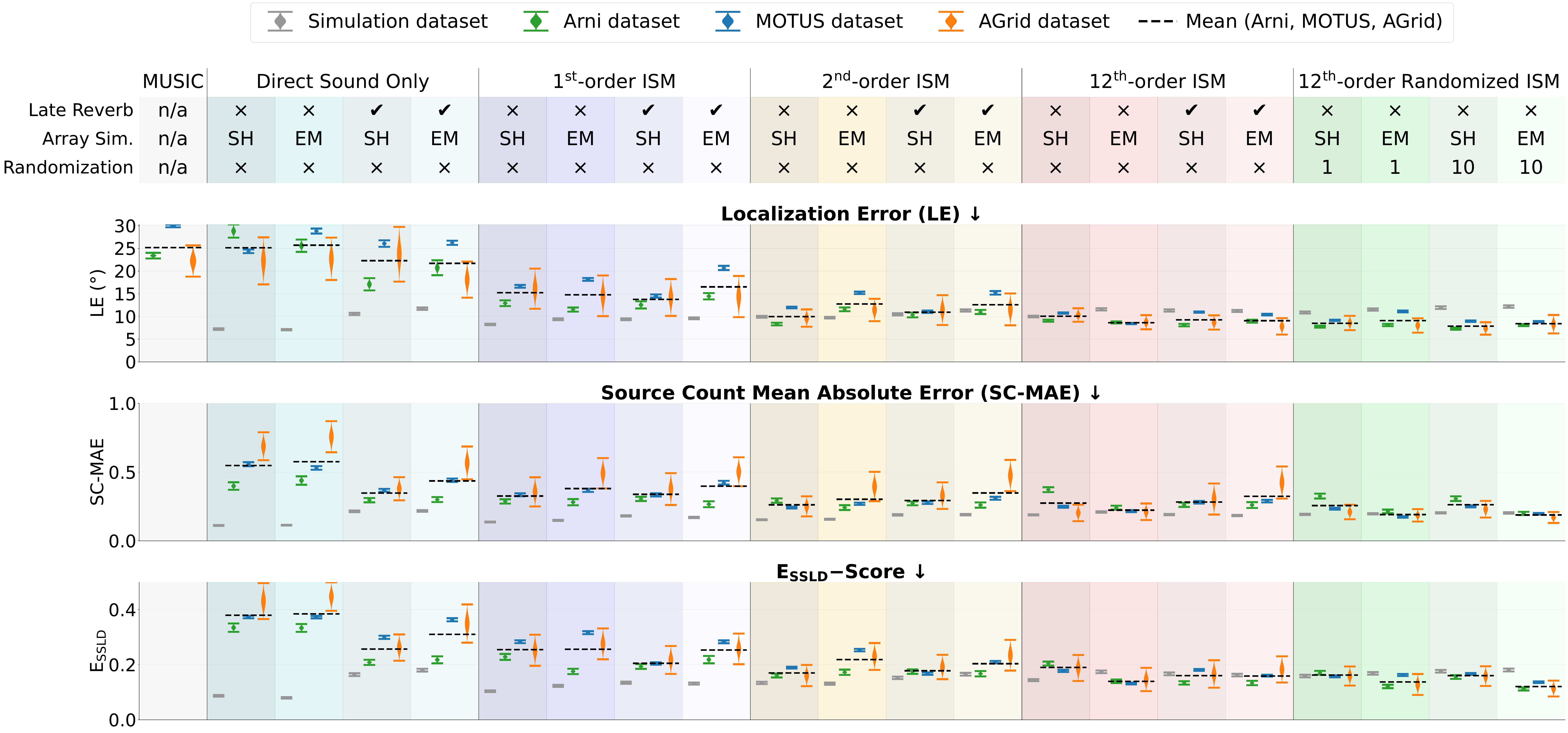}
    \caption{Overview of SSLD performance on a simulation-based and three measurement-based test datasets after training with different acoustic simulators (colored columns). Whiskers indicate 95\,\% confidence intervals.}
\label{fig:overview_results_mic_order_late_random}
\end{figure*}
\section{Results}\label{sec:results}

\subsection{SSLD Performance}
The results are shown in Fig.~\ref{fig:overview_results_mic_order_late_random}. The top row indicates the configuration of the acoustic simulator used to generate the training data, while the bottom three rows report the evaluation metrics. Each column shows model performance on the test split of the corresponding simulated training dataset in gray, together with performance on the three measured test sets. For comparison, the LE of the MUSIC algorithm is shown in the leftmost column.

Overall performance on the simulated test split is similar across training configurations. As expected, models trained under anechoic conditions perform well on the corresponding simulated test set, but performance drops markedly on the measurement-based datasets. Adding diffuse late reverberation to the direct-sound simulations slightly improves generalization, whereas adding first-order image sources yields a substantially larger benefit. Once image sources are included, however, the additional diffuse reverberation tail does not provide a systematic improvement.

Increasing the ISM order further reduces the performance gap on the measurement-based datasets. The largest gains are obtained when moving from anechoic simulation to first- and second-order image sources, while the additional improvement at 12th order is consistent but smaller. Higher ISM orders, however, also reduce performance variability across datasets. The AGrid results show the largest uncertainty, likely due to the smaller dataset size. At low reflection orders, the ideal SH receiver simulation tends to generalize better, whereas the full EM simulation based on measured ATFs performs better at higher ISM orders and yields the best overall SSLD performance.
For the highest-order ISM simulations, LE is lower on the measured datasets than on the simulated test set. In contrast, SC-MAE remains lower on the simulated test set, and further analysis confirmed that the lower LE on the measured datasets is associated with underestimation of the number of active sources.

The four rightmost columns in Fig.~\ref{fig:overview_results_mic_order_late_random} show the results for training sets with different degrees of randomization of the image-source positions. Randomization slightly improves generalization to the measured datasets compared with the corresponding non-randomized simulations, suggesting that perturbing the regular image-source structure of standard shoebox ISM simulations reduces overfitting to idealized geometries. The largest tested randomization range, with image-source displacements of up to $\pm 10$~m, yields the best performance on the measured datasets. Across the conditions with randomization, the EM simulation again achieves the best overall performance.

\subsection{Simulator Complexity}

To establish cost-performance trade-offs, computational simulation cost must be considered. In general, the computational complexity of the image-source method grows exponentially with reflection order, but for shoebox rooms it increases only polynomially, specifically cubically~\cite{kuttruffRoomAcoustics2016}. On a test machine (Apple MacBook Pro with an M1 Pro chip and 16 GB RAM), the mean SRIR simulation time per room, averaged over 1000 rooms, was 0.49 s for reflection order 1, 0.55 s for reflection order 2, and 1.26 s for reflection order 12. Thus, increasing the order from one to two, and possibly to 12, may often be justified. In our setup, randomizing image-source positions did not alter computational cost, as it only required displacing already simulated image sources.

In any case, the image-source computation itself contributed only a small fraction of the overall runtime compared with loading the ATFs in the EM-based simulations. For both direct-sound simulation and 12th-order reflection simulation, the EM-based approach was more than 40 times slower than the corresponding SH-based simulation. Given the relatively modest performance improvements achieved with the EM receiver, this substantial additional cost may not be justified in all applications.

\section{Discussion and Conclusion}\label{sec:conclusion}
This work analyzed the sound source localization and detection performance of a convolutional recurrent neural network trained with geometrical acoustic simulators of varying complexity and evaluated on three measurement-based test datasets. The best SSLD performance was obtained with the highest-order ISM simulations combined with ATF-based microphone array simulation. A further improvement was observed when image-source positions were randomized, suggesting better generalization beyond the limitations of standard shoebox simulations. These findings indicate that highly accurate geometric simulations may not always be necessary and could potentially be replaced by a more computationally efficient randomized approach, such as the statistical model in~\cite{badeauCommonMathematicalFramework2019} or FRAM-RIR~\cite{Luo2024}, which uses random image-source locations and achieved improved generalization on a speech separation task. 

For the best-performing simulators, SSLD performance on the out-of-domain measurement-based test sets was close to that on the within-domain simulated test split, indicating that the benefit of further increasing simulation complexity may be limited. At the same time, the analysis of computational simulation cost revealed a clear complexity-performance trade-off. Higher image-source orders and additional microphone array simulation substantially increased simulation cost while yielding only modest performance gains. Overall, the results indicate how the simulation complexity-performance trade-off can be exploited in future geometrical-acoustic simulation for data-driven SSLD and show that image-source randomization may offer a practical way to simplify geometrical acoustic simulation while improving generalization.

\section{Acknowledgements}

Fabian Staub received support from an Erasmus+ traineeship grant from the European Union. Thomas Deppisch was supported by Meta Reality Labs through the research project MaMeLi. The authors acknowledge the use of HUCE infrastructure of Aalto School of Electrical Engineering and the computational resources provided by the Aalto Science-IT project.

\bibliographystyle{IEEEbib}
\bibliography{references,references_tommi}

\end{document}